\documentclass[conference, 10pt]{IEEEtran}

\usepackage{bm}
\usepackage[usenames,dvipsnames]{xcolor}
\usepackage{srcltx}
\usepackage{psfrag}
\usepackage{epsfig}
\usepackage{graphicx}
\usepackage{epstopdf}
\usepackage{color}
\usepackage[tbtags,sumlimits,nointlimits,reqno]{amsmath}
\usepackage{amssymb}
\usepackage{color}
\usepackage{float}
\usepackage{gensymb}
\usepackage{subfig}
\usepackage{cite}

\begin{document}

\title{Study of the Angular Spectrum of a Bianisotropic Refractive Metasurface at a Dielectric Interface}

\author{\IEEEauthorblockN{Guillaume~Lavigne and Christophe~Caloz}
\IEEEauthorblockA{Department of Electrical Engineering\\ Polytechnique Montr\'{e}al\\
 Montr\'{e}al, Qu\'{e}bec, Canada
}
\and
\IEEEauthorblockN{Karim Achouri}
\IEEEauthorblockA{Nanophotonics and Metrology Laboratory \\ \'{E}cole Polytechnique F\'{e}d\'{e}rale de Lausanne\\
 Lausanne, Vaud, Switzerland\\
}
}
\maketitle
\begin{abstract}
We present an initial study of the angular spectrum of a bianisotropic refractive metasurface at an interface between two dielectric media. In this study, we report on the existence of three distinct angular regions: a)~a rotated transmission cone, b)~a modified total internal reflection region, and c)~a new total retro-reflection region.
\end{abstract}

\section{Introduction}\label{sec:intro}

Metasurfaces have recently received considerable interest due to their unprecedented ability to control electromagnetic fields and resulting virtually unlimited applications~\cite{glybovski2016metasurfaces}.

One of the most fundamental operations enabled by metasurfaces is generalized refraction~\cite{yu2011light}, which has recently been demonstrated to require bianisotropy~\cite{asadchy2016perfect,epstein2016arbitrary,lavigne2017refraction} in order to avoid undesired diffraction orders.

Here, we present a preliminary study of the angular spectrum of such a bianisotropic refractive metasurface placed at the interface between two dielectric media, and reveal the existence of three regions.

\section{Metasurface Synthesis}\label{sec:synthesis}

We use a susceptiblity-GSTCs (Generalized Sheet Transition Conditions) metasurface synthesis method~\cite{achouri2014general} to design the bianisotropic refractive metasurface. Assuming a purely transverse metasurface, the GSTCs read
\begin{subequations}\label{eq:GSTC}
\begin{equation}
\hat{z} \times \Delta\mathbf{H} = j \omega \epsilon_0 \overline{\overline{ \chi}}_\text{ee} \cdot \mathbf{E}_\text{av} +  j k_0 \overline{\overline{ \chi}}_\text{em}  \cdot\mathbf{H}_\text{av} ,
\end{equation}
\begin{equation}
\Delta \mathbf{E} \times \hat{z}   = j \omega \mu_0 \overline{\overline{ \chi}}_\text{mm}\cdot \mathbf{H}_\text{av} +  j k_0 \overline{\overline{ \chi}}_\text{me} \cdot  \mathbf{E}_\text{av} ,
\end{equation}
\end{subequations}
where the $\Delta$ symbol and the 'av' subscript represent the differences and averages of the tangential fields on both sides of the metasurface, and $\overline{\overline{ \chi}}_\text{ee}, \overline{\overline{ \chi}}_\text{em}, \overline{\overline{ \chi}}_\text{me}, \overline{\overline{ \chi}}_\text{mm}$ are the bianisotropic surface susceptibility tensors describing the metasurface. This method consists in the following three steps: 1)~specifying the total fields on both sides of the metasurface, 2)~calculating the difference and average tangential fields, and 3)~inserting these fields into Eqs.~\eqref{eq:GSTC} to obtain the susceptibility functions.

\begin{figure}[h]
\centering
\includegraphics[width=0.8\columnwidth]{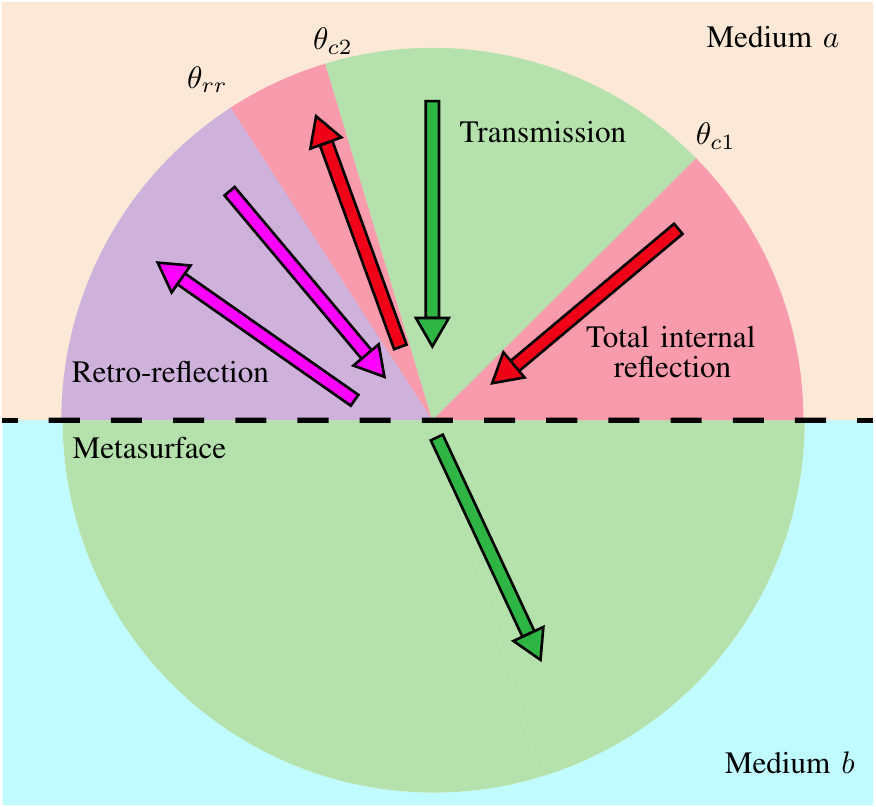}
\caption{Angular regions with different scattering responses for an interface between two dielectric media [(a) and (b)] hosting a bianisotropic refractive metasurface.}\label{fig:regions}
\end{figure}

We consider a metasurface placed at the interface between two dielectric media, denoted $a$ and $b$, with parameters $\epsilon_a$, $\eta_a$, $k_a$ and $\epsilon_b$, $\eta_b$ and $k_b$, respectively, as shown in Fig.~\ref{fig:regions}. In this system, a plane wave incident from medium $a$ at an angle $\theta_a$ is refracted, without any diffraction orders, into medium $b$ at an arbitrary angle $\theta_b$.

This problem was previously solved for a general bianisotropic refractive metasurface for TM polarization in~\cite{lavigne2017refraction}. The corresponding susceptibility functions, derived in that work, are
\vspace{-3mm}
\begin{subequations}\label{eq:biani_X}
\begin{equation}
\chi^{xx}_\text{ee} =
\frac{-4 k_a k_b  T  \sin (\alpha x)}{\epsilon_0 \omega \left(T  \beta \cos (\alpha x)+\eta_a k_b k_{az}+\eta_b k_a k_{bz} T ^2\right)},
\end{equation}
\begin{equation}
\chi^{xy}_\text{em} =
\frac{2 j \left(T  \gamma \cos (\alpha x)-\eta_b k_b k_{az}+\eta_a k_a k_{bz} T ^2\right)}{k_0 \left(T  \beta \cos (\alpha x)+\eta_a k_b k_{az}+\eta_b k_a k_{bz} T ^2\right)},
\end{equation}
\begin{equation}
\chi^{yx}_\text{me} =
\frac{2 j \left(T  \gamma \cos (\alpha x)-\eta_b k_b k_{az}+\eta_a k_a k_{bz} T ^2\right)}{k_0 \left(T  \beta \cos (\alpha x)+\eta_a k_b k_{az}+\eta_b k_a k_{bz} T ^2\right)},
\end{equation}
\begin{equation}
\chi^{yy}_\text{mm} =
\frac{-4 \eta_a \eta_b k_{az} k_{bz}  T  \sin (\alpha x)}{\mu_0 \omega \left(T  \beta \cos (\alpha x)+\eta_a k_b k_{az}+\eta_b k_a k_{bz} T ^2\right)},
\end{equation}
\end{subequations}
where $\alpha=k_{ax}-k_{bx}$, $\beta = \eta_a k_b k_{az}+\eta_b k_a k_{bz}$ and $\gamma = \eta_b k_a k_{bz}-\eta_a k_b k_{az}$. The transmission coefficient $T =\sqrt{(\eta_b\cos\theta_a)/(\eta_a\cos\theta_b)}$ is obtained by enforcing power conservation across the metasurface such that all incident power is transmitted.

\begin{figure*}[h!t]
\centering
\subfloat[Transmission region: excitation at $\theta_a=0$ (specified incidence) with expected diffraction-free refraction at $\theta_b=25^\circ$.]{{\includegraphics[width=0.3\textwidth]{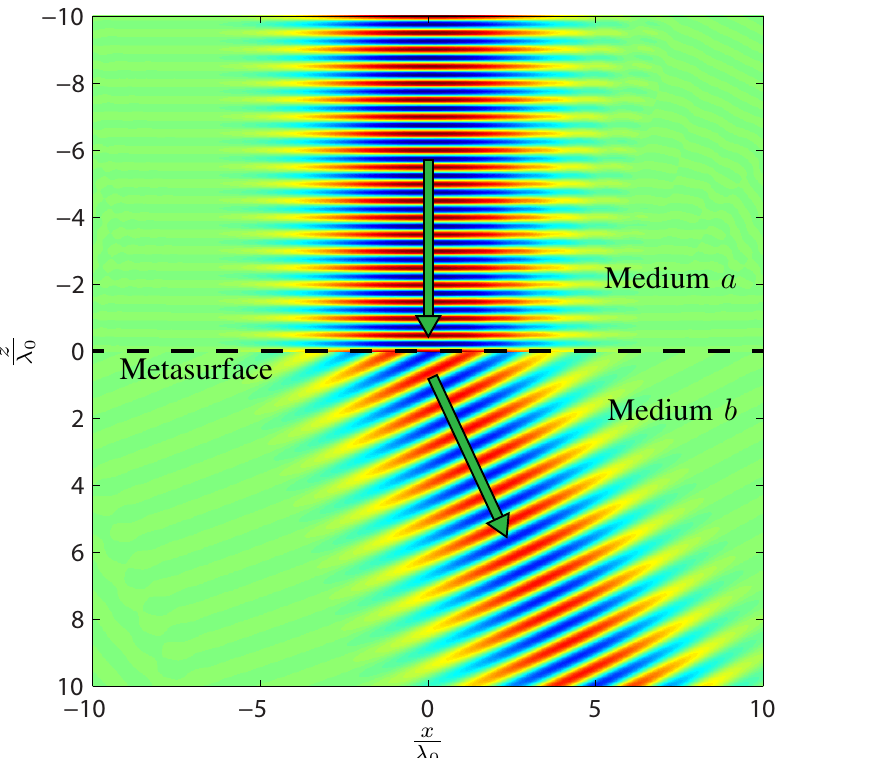}}}
\qquad
\subfloat[Total internal reflection region: excitation at $\theta_a=50^\circ$, with emergence of diffraction orders.]{{\includegraphics[width=0.3\textwidth]{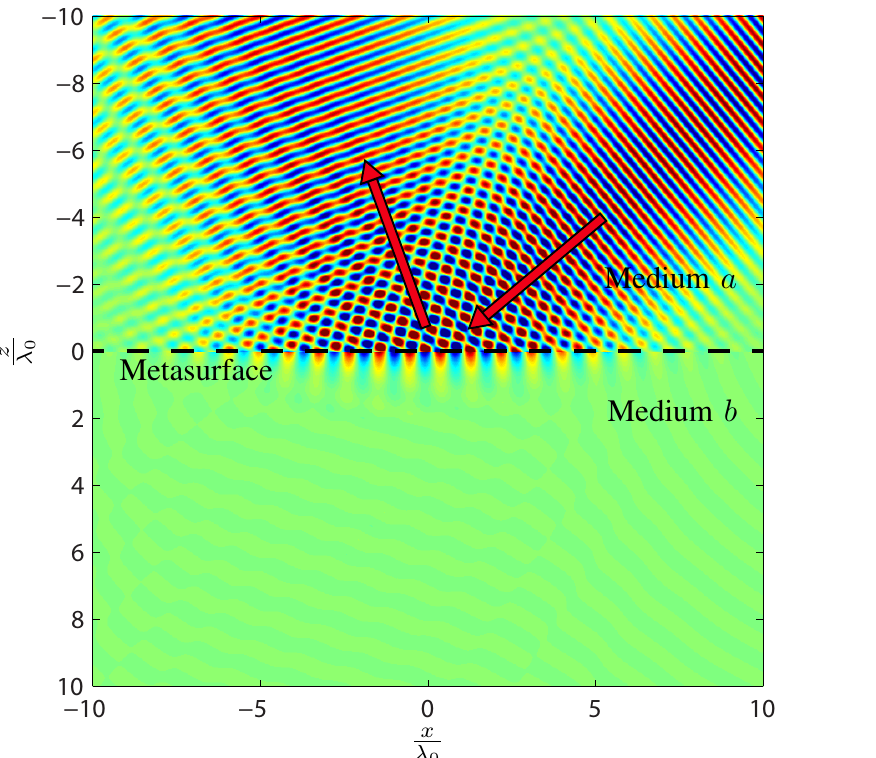}}}
\qquad
\subfloat[Total retro-reflection region: excitation at $\theta_a=-40^\circ$, and reflection at $\theta=-55^\circ$.]{{\includegraphics[width=0.3\textwidth]{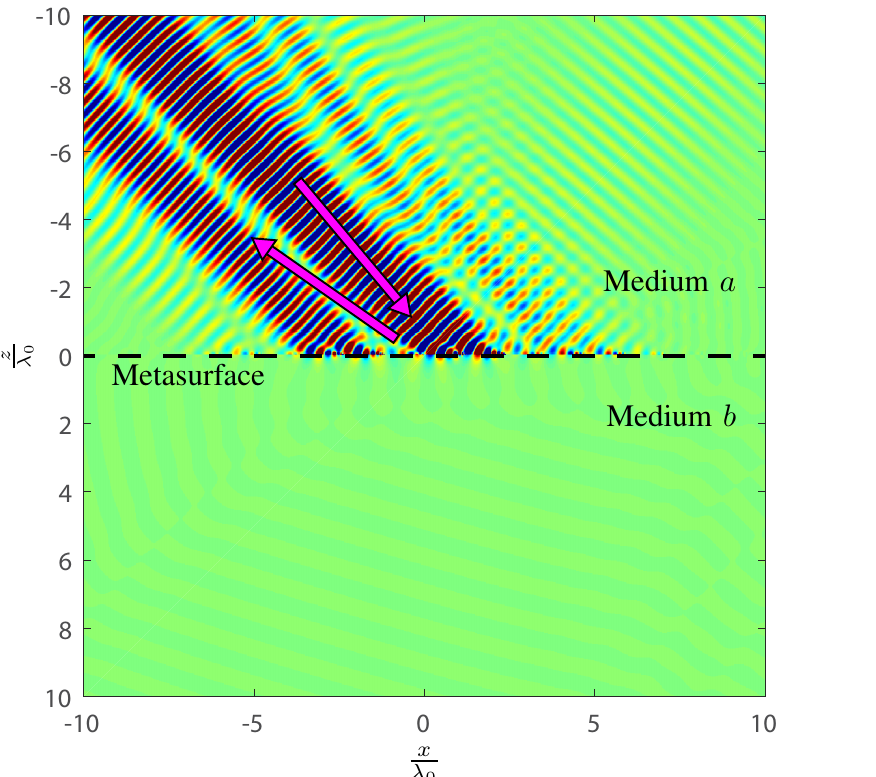}}}
\caption{Full-wave GSTC-FDFD simulated~\cite{vahabzadeh2016simulation} response of the refractive bianisotropic metasurface at the interface between a dielectric medium with $\epsilon_{\text{r}a}=4$ and air ($\epsilon_{\text{r}b}=1$) to a Gaussian beam, with metasurface interface \emph{synthesized for} incidence at $\theta_a=0$ and refraction at $\theta_b=25^\circ$.}
\label{fig:simul}
\end{figure*}

These susceptibilities correspond to a periodic bianisotropic metasurface, which is passive, lossless and reciprocal. The metasurface can be tailored to accommodate arbitrary media surrounding it since its susceptibilities are a function of $\eta_{(a,b)}$ and $k_{(a,b)}$. It adds the momentum $K_x = k_{bx}-k_{ax}$ to the wave vector of the incident wave $k_{ax}$, which is akin to the phase gradient in the generalized laws refraction and reflection~\cite{yu2011light}.

\section{Angular Spectrum Study}\label{sec:ang_study}

We shall see that the metasurface-interface admits three distinct \emph{asymmetric} (due to the metasurface momentum) angular regions, as indicated in Fig.~\ref{fig:regions}: a)~a \emph{transmission region}, corresponding to a rotated escape cone, b)~a \emph{total internal reflection region}, very different from that of the conventional interface, and c)~a \emph{total retro-reflection region}, non existing in the conventional interface.

We consider here a bianisotropic metasurface interface synthesized to refract a plane wave incident at normal incidence ($\theta_a=0^\circ$) in a dielectric medium $a$ with $\epsilon_{\text{r}a}=4$ into a plane wave propagating at an angle $\theta_b = 25^\circ$ in air (medium $b$). We shall now present GSTC-FDFD simulations~\cite{vahabzadeh2016simulation} for the three aforementioned angular regions.

A simulation for the transmission region is shown in Fig.~\ref{fig:simul}(a), specifically for the excitation corresponding to the synthesis specification. As in the case of a phase-gradient metasurface~\cite{yu2011light}, this transmission region is delimited by the two critical angles
\vspace{-1.5mm}
\begin{equation}\label{eq:critical_angles}
\theta_{c1} = \arcsin\left(\frac{K_{x}+n_{b}}{n_a}\right),
\;
\theta_{c2} = \arcsin\left(\frac{K_{x}-n_{b}}{n_a}\right),
\end{equation}
which derives from the momentum added by the metasurface to the wavevector, $K_x = k_{bx}-k_{ax}$, and correspond to $\theta_{c1}=45.3^\circ$ and $\theta_{c2}=-16.8^\circ$ in our example.

Figure~\ref{fig:simul}(b) shows a simulation for the total internal reflection region. The scattering in this region is similar to that of a conventional interface, except for its asymmetry. This region is composed of two sectors, delimited by delimited by $\theta_{c1}$ and $\theta=+\pi/2$ and by $\theta_{c2}$ and an angle $\theta_\text{rr}$, that is specific to the \emph{metasurface} interface.

The angle $\theta_\text{rr}$ corresponds to the limit where the incident wave would be reflected at $\theta=+\pi/2$ and beyond which the wave also becomes evanescent in the incident medium. Its value is given by
\vspace{-1.5mm}
\begin{equation}\label{eq:angle_RR}
\theta_\text{rr}= \arcsin\left(\frac{2K_{x}}{n_a}-1\right),
\end{equation}
which numerically corresponds here to $\theta_\text{rr}=-35.3^\circ$. We observe by full-wave simulation, without explanation so far, that this phenomenon results total retro-reflection beyond $\theta_\text{rr}$, as shown in Fig.~\ref{fig:simul}(c). This is an interesting and potentially useful new property of this bianisotropic metasurface interface.

\vspace{-1.5mm}
\section{Conclusion}\label{sec:conclusion}

We presented an initial study of the angular spectrum of a bianisotropic refractive metasurface interface at an interface between two dielectric media, and reported on the existence of three scattering regions. This study provides an insight in the extra effects achievable at an interface hosting a metasurface.

\vspace{-1.5mm}
\bibliographystyle{IEEEtran}
\bibliography{LIB}

\end{document}